# Spatial data science: Closing the human-spatial computing-environment loop
Benjamin Adams, University of Canterbury, New Zealand

Over the last decade, the term 'spatial computing' has grown to have two different, though not entirely unrelated, definitions. The first definition of spatial computing stems from industry, where it refers primarily to new kinds of augmented, virtual, mixed-reality, and natural user interface technologies.  A second definition coming out of academia takes a broader perspective that includes active research in geographic information science as well as the aforementioned novel UI technologies (Shekar et al. 2016). Both senses reflect an ongoing shift toward increased interaction with computing interfaces and sensors embedded in the environment and how the use of these technologies influence how we behave and make sense of and even change the world we live in.

Regardless of the definition, research in spatial computing is humming along nicely without the need to identify new research agendas or new labels for communities of researchers. However, as a field of research, it could be helpful to view *spatial data science* as the glue that coheres spatial computing with problem-solving and learning in the real world into a more holistic discipline. Thus, spatial data science is expressly concerned with problems that involve not only computational modeling and representation of the environment (e.g., using GIS technologies) but also people—what their concerns are about the environment and how they behave in space and alter those spaces, as well as how the use of spatial computing technologies (broadly conceived) influences each of these. Starting from this premise, I propose that a systems theory approach to analyzing the complex feedback relationships between people, the environment, and spatial computing information technologies can help to clarify the research challenges that spatial data science is uniquely qualified to address and provide a theoretical basis for understanding what constitutes spatial data science as a field of research.

The science of systems has been successfully applied to understand processes studied in diverse disciplines from the social sciences and cybernetics to ecology and environmental science (Banathy 2013, Alberti et al. 2011). In systems thinking some of the fundamental concerns include feedback loops and non-linear dynamics, the controllability and observability of systems, and notions such as equifinality, which states that there are many different routes that lead to the same system state. We do not yet have a good understanding of systems that involve spatial data.

Figure 1, left shows the duality of space-time and information in a spatial computer as defined in Beal et al. (2013). This is an example of a coupled information-environment system, where the use of a spatial computer can perform four operations that affect the internal state of the information system and the physical space-time that the spatial computer spans: measuring space-time, manipulating space-time, computing patterns, and physical evolution. Although originally conceived as a model to represent changes in the space-time characteristics of a spatial computer per se, a similar approach could be used to model the wider physical environment and the physical evolution of that environmental system in response to the manipulations of a spatial computer (e.g., a UAV that trims trees). Human ecology (a.k.a. coupled human environment systems) is another example of systems theory applied to the complex relationships between humans and the natural world that looks at reciprocal interactions between humans and the environment. Figure 1, right shows a basic schematic of this kind of system (Alberti et al. 2011).

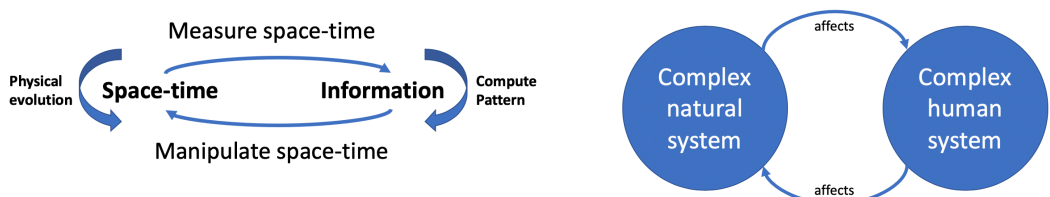

*Figure 1. System model of a spatial computer from Beal et al. 2013, left. Coupled human-environment system, right.*

**Human-spatial computing-environment systems**

The shift to real-time generation of massive amounts of data coupled with embedded and ubiquitous spatial computing means that direct and complex feedback loops now exist between dynamic human, environment, and spatial computing systems. An example of a feedback effect in a *human-spatial computing-environment system* is the Instagram effect in tourism where a point-of-interest becomes popular through use of social media, which then changes future tourist behavior and possibly the sustainability of an entire tourism system that depends on complex relationships between environmental and social components. Another example is the use of spatial computing for learning about the physical world and the impacts of such learning on human behavior and the environment. Speculative technologies such as the AR cloud, which would create a dynamic sub-centimeter map of the world using computer vision, would represent a human-spatial computing-environment system of unparalleled complexity.

**Implications for spatial data science**

While not every spatial data science research project need incorporate full *human-spatial computing-environment systems thinking*, it can help drive how we organize some of our research priorities. Similar to arguments made early on in the development of GIScience regarding spatial data handling, it is certainly not enough for spatial data science to resolve to a collection of methods and tools for working with spatial data. A human-spatial computing-environment systems approach can encourage the field to engage and draw from the problems and concerns of researchers outside of the existing GIScience community. Connections between humans and the environment are as important as any other relationship in the system. Thus, we need strong connections to HCI research and deep understanding of the problems that "domain" researchers are engaged with at the human-environment interface. On a meta-level, a challenge is to understand the observability of such a system—a topic that has been alluded to in recent discussions around geographic information observatories (Adams & Gahegan 2016).

In this workshop I am keen to explore some of the key relationships (coupled, ternary, etc.) in human-spatial computing-environment systems and the implications of those relationships, such as feedbacks and emergent properties that might arise. What are the key components of such a system? Does a systems approach like this incentivize the prioritization of dynamic process-based modeling in spatial data science (and thus help us avoid overly simplistic static spatial models)? How does a system thinking approach help us to identify new research problems in spatial data science?